\documentclass[12pt]{article}
\usepackage{latexsym}
\usepackage{amsmath}
\usepackage{indentfirst}
\usepackage{cite} 

\long\def\ca#1\cb{} 

\newcommand{\ad}{^\dagger }

\newcommand{\AND}{\mbox{\small AND}}

\newcommand{\becs}{\begin{cases}}
\newcommand{\bem}{\begin{matrix}}

\newcommand{\dya}[1]{|#1\rangle\langle#1|}

\newcommand{\encs}{\end{cases}}
\newcommand{\enm}{\end{matrix}}

\newcommand{\inp}[1]{\langle#1|#1\rangle }
\newcommand{\inpd}[2]{\langle#1|#2\rangle }
\newcommand{\ket}[1]{|#1\rangle }

\newcommand{\lra}{\leftrightarrow }

\newcommand{\mted}[3]{\langle#1|#2|#3\rangle }

\newcommand{\ot}{\otimes }

\newcommand{\ra}{\rightarrow }
\newcommand{\Ra}{\Rightarrow }

\newcommand{\st}{\sqrt{2}}

\newcommand{\AC}{{\mathcal A}}
\newcommand{\BC}{{\mathcal B}}

\newcommand{\al}{\alpha }
\newcommand{\bt}{\beta }

\newcommand{\dl}{\delta }
\newcommand{\Dl}{\Delta }
\newcommand{\ep}{\epsilon}

\renewcommand{\th}{\theta } 

\newcommand{\lm}{\lambda }

\newcommand{\sg}{\sigma }

\def\outl#1{\par{\medskip\noindent\hspace*{0.1cm}\bf
      \mathversion{bold}#1\mathversion{normal}\smallskip} }
   \def\xa{} \def\xb{}  

 \def\outl#1{}\def\xa{}\def\xb{}

\ca
 \def\outl#1{\par{\medskip\noindent\hspace*{.5cm}\bf
      \mathversion{bold}#1\mathversion{normal}\smallskip} }
 \long\def\xa#1\xb{} 
  
\cb

\begin{document}

\title{Multitime Quantum Communication: Interesting But Not Counterfactual}
\author{Robert B. Griffiths\thanks{Electronic address: rgrif@cmu.edu}\\
Department of Physics\\
Carnegie Mellon University\\
Pittsburgh, PA 15213}

\date{Version of 24 June 2023}
\maketitle

\begin{abstract}
  A protocol for transmission of information between two parties introduced by
  Salih et al., {\em Phys. Rev. Lett.} 110 (2013) 170502 (hereafter SLAZ),
  involves sending quantum amplitude back and forth through a quantum channel
  in a series of steps, rather than simply sending a signal in one direction.
  The authors claimed that their protocol was ``counterfactual'' in the sense
  that while a quantum channel is needed to connect the parties, its actual
  usage becomes vanishingly small in the asymptotic limit as the number of
  steps tends to infinity. Here we show that this claim is incorrect because it
  uses probabilistic reasoning that is not valid at intermediate times in the
  presence of quantum interference. When ill-defined probabilities are replaced
  with a well-defined measure of channel usage here called ``Cost'', equal to
  the absolute square of the amplitude sent through the channel, the total Cost
  does not go to zero in the asymptotic limit of a large number of steps, but
  is bounded below by a rigorous inequality. A detailed analysis shows that
  this bound is satisfied in the SLAZ protocol. The analysis leading to the
  bound uses the fact that the Gram matrix formed by inner products of a
  collection of pure quantum states is additive over Hilbert subspaces and
  invariant under unitary time transformations. Its off-diagonal elements,
  which in general are not positive, play a significant role in the formal
  argument as well as providing a somewhat strange way of visualizing the
  transfer of information.
\end{abstract}

\tableofcontents

\section{Introduction \label{sct1}}

\xb
\outl{OVERVIEW OF PAPER}
\xa

\xb 
\outl{SLAZ protocol: Information transmitted via amplitude moving
  back-and-forth thru channel in a series of unitary time steps} 
\xa

The motivation for this paper is a scheme for the transmission of quantum
informatiom introduced by Salih et.\ al \cite{Slao13} with the title ``Protocol
for direct counterfactual quantum communication'', and referred to hereafter as
SLAZ, the initials of the authors. One ordinarily thinks of the transmission of
information as sending a signal through a channel from sender to receiver.
However the idea in SLAZ is that information can be sent from Bob to Alice if
the quantum particle used to carry the information starts off in Alice's
domain, and a part of its quantum amplitude is sent to Bob through a quantum
channel. Bob modifies this is some way before sending (or possibly not sending)
it back to Alice, depending on the signal he wants to send. Alice then employs
what Bob has returned to begin a second round of sending amplitude to Bob, who
again modifies it before returning it, and so forth. This back-and-forth motion
can continue for a large number of rounds until the information that Bob is
sending has arrived in Alice's domain, where she can carry out a measurement or
perhaps perform additional processing. A key feature of protocols of this type
is that all the intermediate steps can be represented by purely unitary time
evolution, with intermediate time measurements, if any replaced by
unitaries---a process of purification.

\xb
\outl{Use of 'amplitude' rather than 'particle', which is associated with 
location in space. Qm properties $\lra$ projectors; wave $[\psi]$ does not
commute with position projector. Double slit}
\xa

The use of \emph{amplitude} rather than \emph{particle} in the previous
paragraph is intentional, because the state of the photon or other particle is
in general a coherent superposition of parts associated with different spatial
locations: Alice's domain, Bob's domain, and the channel connecting them. One
generally thinks of a particle as something with a spatial location, but in
quantum mechanics one cannot simultaneously ascribe particle and wave
properties to the same entity at the same time because of wave-particle
duality. In Hilbert-space quantum mechanics physical properties, such as
location in space, are represented by \emph{projectors} (Sec.~III.5 of
\cite{vNmn32b}), and when a projector representing a wave, think of
$\dya{\psi}$, does not commute with a projector specifying a spatial location,
ignoring this fact can rapidly lead to paradoxes. The double-slit paradox is an
example: when a coherent wave passes through the slit system one cannot say
through which slit the particle passed.

\xb
\outl{'Counterfactual' in SLAZ: Channel essential, but use in each step can be
  made arbitrarily small}
\xa

The term ``counterfactual'' in the original SLAZ paper has the following
significance. A quantum channel connecting the communicating parties is
essential: this is not a case of mysterious nonlocal influences of the sort
which are sometimes invoked to explain quantum violations of Bell inequalities.
However, if the number of steps in an SLAZ protocol is sufficiently large, the
magnitude of the amplitude sent through the channel in each step can be made
very small, and vanishes in the limit as the number of steps tends to infinity.

\xb
\outl{Literature generated by SLAZ. Papers with bibliographies. Sec.~\ref{sct5}
has some comments}
\xa

A similar claim of counterfactuality has been made in much of the rather
substantial literature motivated by the original SLAZ publication, which
contains various modifications and extensions of the original protocol. There
have also been criticisms of these counterfactual claims, and (of course)
replies to criticisms. The Conclusion, Sec.~\ref{sct5}, of the present paper
contains a few remarks about how its results apply to some of these
publications, but a review, much less a detailed discussion, lies far outside
its scope. The interested reader is referred to the extensive bibliographies
found in \cite{HnLR21,Liao22}.

\xb
\outl{AIM \& CONTENTS OF PRESENT PAPER}
\xa

\xb
\outl{Aim: Study CHANNEL USAGE in SLAZ \& related}
\xa

\xb
\outl{Replace ill-defined probs at intermediate times with COST: Absolute
  square of amplitude sent thru channel.}
\xa

\xb
\outl{Sec. 2 discusses use of Cost in case where careless invocation of probs
  is related to paradoxes, and shows that Cost is a reasonable measure in
multitime one-way communication protocols}
\xa

The aim of the present paper is to study the use of quantum channels in
protocols of the SLAZ type, in particular the sense in which this usage is or
is not counterfactual. To this end a technical term, \emph{Cost}, the absolute
square of the amplitude through the channel in a particular step in the
protocol, is used for reasons discussed in Sec. II, as a useful substitute for
"probability", which in a quantum context is often ill-defined. The example of
multiple channels in parallel, which few would want to claim are
counterfactual, serves as an introduction to how information can be sent
through a single channel in a single direction at multiple times, in a process
in which all of the intermediate steps are represented by unitary maps.

\xb
\outl{Sec. III. Main mathematical results}
\xa

The main mathematical results of this paper are in Sec.~\ref{sct3}: Gram
matrices and some of their properties are discussed in Sec.~\ref{sbct3a}, while
Sec.~\ref{sbct3b} gives the basic structure of simple two-way multiple time
protocols. Section~\ref{sbct3c} considers simple schemes for transmitting one
classical bit, while the rigorous lower bound that undermines various
counterfactual claims is the topic of Sec.~\ref{sbct3d}.

\xb \outl{Original SLAZ protocol studied in detail in Sec. 4. In particular the
  total Cost evaluated for transmitting each of the bits. Remarks on Costs of
  $\lm=0$ (reflection) and $\lm=1$ (absorption) are very different, but their
  product satisfies rigorous bound of Sec.~\ref{sbct3d}} \xa

\xb
\outl{Mistaken SLAZ ctfl claim: Measure of channel usage which would not work
  in a Cl contest; probs that are invalid in a Qm context}
\xa

The original SLAZ protocol is studied in detail in Sec.~\ref{sct4}. In
particular the total Cost of transmitting a classical bit $\lm=0$, in which Bob
reflects the amplitude back to Alice, and for transmitting $\lm=1$, in which he
absorbs rather than returns it, are evaluated explicitly. It turns out that in
the asymptotic limit the $\lm=1$ Cost is miniscule, but that for $\lm=0$ is
enormous, while the product of the two remains finite and satisfies the
rigorous bound in Sec.~\ref{sbct3d}. The mistaken claim that the SLAZ protocol
is counterfactual results from two errors: a concept of channel use which would
be questionable even for a classical stochastic process, and an improper use
of probabilities in a way that violates quantum principles.

\xb
\outl{Concluding Sec. 5. Some comments on literature engendered by SLAZ.
  Summary of results of this paper. Suggestions for directions of future
research.}
\xa

The concluding Sec.~\ref{sct5} has a summary of the main results of this paper,
a few comments on some parts of the literature related to SLAZ, and some
suggestions for future directions of research. This author believes that
protocols of the SLAZ type are quite interesting, deserve further exploration,
and might contribute to useful ways of studying multipartite and multitime
transmission of quantum information, as in quantum networks. And that such
studies would prove more fruitful in the absence of claims of
counterfactuality.


\section{ One-Way Protocols \label{sct2}}

\subsection{ Multiple Channels in Parallel \label{sbct2a}}
\xb
\outl{Photon carrying info could be sent thru N channels in parallel.}
\xa

\xb
\outl{$c_n=$ amplitude in channel $n$;
 $q_n := |c_n|^2$; $Q:=\sum_n q_n$ }
\xa

\xb \outl{Why $q_n$ should NOT be called a PROBABILITY. Probs at intermediate
  times not defined. Measurement does not help} \xa

Think of quantum information as the information carried by a photon as it
passes through a quantum channel, such as an optical fiber. The information
could be encoded in its polarization. Rather than using a single channel, one
could imagine sending the photon as a superposition state through a set of $N$
channels in parallel, using a collection of beamsplitters to divide up the
initial amplitude among the different channels, and a corresponding collection
to later recombine them.
Let us suppose that the normalized $\ket{\Phi}$ that represents the photon
at some intermediate time is a coherent superposition of amplitudes
\begin{equation}
 \ket{\Phi} = \sum_{n=1}^N c_n \ket{\phi_n}
\label{eqn1}
\end{equation}
associated with the individual channels, labeled by $n$.  Define the
\emph{Cost} $q_n$ associated with the use of channel $n$, and the \emph{total
  Cost} $Q$ for the channel system as:
\begin{equation}
 q_n := |c_n|^2,\quad Q := \sum_{n=1}^N q_n.
\label{eqn2}
\end{equation}
If the $\ket{\phi_n}$ and $\ket{\Phi}$ are normalized, $Q$ is equal to 1, so
one might identify $q_n$ with the \emph{probability} that the photon is in
channel $n$. But what does that mean? In standard (textbook) quantum mechanics
probability refers to the outcome of a measurement, but a measurement carried
out at an intermediate time, when the quantum state is a coherent superposition
over various locations, can alter what occurs later, and hence it is dangerous
to associate such a probability with a situation in which a measurement does
\emph{not} take place.

\xb
\outl{vN: Qm property $\lra$ Hilbert space projector. 
 Spin-half Sz, Sx illustration}
\xa

\xb
\outl{Standard probabilities require PDI as sample space}
\xa

Another way of viewing this difficulty is to recall that von Neumann
(Sec.~III.5 of \cite{vNmn32b}) identified quantum \emph{physical
  properties}---which in classical physics are associated with sets of points in
the classical phase space---with \emph{projectors}, self-adjoint idempotent
operators, $P=P\ad=P^2$, on the quantum Hilbert
space. For example, in the case of a spin-half particle the
projectors 
\begin{equation}
 P = (I-\sg_z)/2 \quad R= (I+\sg_x)/2,
\label{eqn3}
\end{equation}
where I is the identity and $\sg_z$ and $\sg_x$ are Pauli operators, represent
the properties $S_z=-\hbar/2$ and $S_x=+\hbar/2$, respectively. In general, if
two projectors $P$ and $R$ commute their product $PR=RP$ represents the
property $P\ \AND\ R$. But if they do not commute, neither $PR$ nor $RP$ is a
projector, and so neither represents a quantum property. In some sense
noncommutation is the very essence of quantum mechanics; it is what
distinguishes it from classical physics.
The use of standard (Kolmogorov) probabilities requires a \emph{sample space}
of mutually-exclusive possibilities, one and only one of which occurs in a
particular run of an experiment. In quantum theory such a sample space is a
collection of mutually orthogonal projectors that sum to the identity, a
\emph{projective decomposition of the identity}. For example, $R$ and $I-R$ in
\eqref{eqn3} in the case of spin half; see \eqref{eqn7} below for the
general definition. In quantum mechanics there are often many possible sample
spaces that one might be interested in, and carelessly combining incompatible
spaces---some projectors in one do not commute with projectors in the
other---inevitably leads to paradoxes rather than physical understanding.

\xb
\outl{Present application: $[\Phi]$ and $[n]$ do not commute}
\xa

In the present context the dyad $\dya{\Phi}$ is a projector that does not
commute with any of the projectors $\dya{\phi_n}$ for which $c_n$ is nonzero,
and thus it is meaningless to assign a probability to the latter in a situation
where the coherent superposition $\ket{\Phi}$ will later be transformed by the
final beamsplitters into the original state that entered the channel system.
For example, in a double-slit experiment in which the amplitudes from the two
slits combine coherently to produce interference, it makes no sense to talk
about the probability that the photon previously passed through one slit rather
than the other. The two costs are well-defined: they are simply the absolute
squares of the two amplitudes. But attempting to measure which slit the
particle passes through in order to define a probability will destroy the
interference pattern.

\subsection{ One Channel Used Multiple Times \label{sbct2b}}

\xb
\outl{Photon can be in three locations $A$ (Alice), $B$ (Bob), $C$ (channel)}
\xa

\xb
\outl{Projectors $A$, $B$, $C$ mutually orthogonal, sum to identity $I$}
\xa

\xb
\outl{Additional degrees of freedom, e.g. polarization, and be represented by
tensor product. $A$ in this case means $A\ot I_f$}
\xa

The possible advantages, if any, of using many channels in parallel can also be
realized by employing a \emph{single} channel and sending quantum amplitude
through it at a \emph{succession} of times; this is what makes protocols of the
SLAZ type of some interest. Let us suppose that information is being sent from
Bob to Alice. One can think of the photon at a particular
time as being in a coherent superposition of amplitudes in three
different physical locations: Alice's domain $A$, Bob's domain $B$, and the
channel $C$ connecting them. The same symbols can be used for the
Hilbert-space projectors associated with these locations, thus operators which
are self-adjoint and idempotent, $A=A\ad=A^2$, and mutually orthogonal,
$AB=BC=AC=0$. They sum to the identity
\begin{equation}
 A + C + B =I
\label{eqn4}
\end{equation}
and hence form a \emph{projective decomposition of the identity}---see the
general definition in \eqref{eqn7} below. A projective decomposition of the
identity is the quantum counterpart of the sample space of mutually exclusive
possibilities essential for using standard (Kolmogorov) probability theory in
the case of a quantum system. Note that $A$, $B$, and $C$ are \emph{subspaces}
of a single Hilbert space, not \emph{subsystems} represented by a tensor
product. If the quantum particle possesses other degrees of freedom, these
projectors are to be understood using the usual convention as including the
identity operator on these additional degrees of freedom. Thus for a photon,
$A$ means that it is located in Alice's domain, whatever may be its
polarization.

\xb
\outl{Bob sends to Alice info of type $\lm$ using unitary $\BC^\lm$ applied to
  initial $\ket{\psi_0}$. Alice removes it from channel using unitary $\AC$}
\xa

Bob can send a particular type of information $\lm$ to Alice by starting with a
normalized \emph{reference state} $\ket{\psi_0}= B\ket{\psi_0}$, the particle
is somewhere in his domain $B$, and using a unitary transformation $\BC^\lm$
acting on the subspace $B+C$ to place it in a state
\begin{equation}
 \ket{\psi_1^\lm} = C\ket{\psi_1^\lm}=\BC^\lm\ket{\psi_0}, 
\label{eqn5}
\end{equation}
in the channel, at which point it travels through the channel to Alice. As the
channel has no effect except to transmit the particle from one end to the
other, we simplify the discussion (here and later) by using the same symbol for
the ket that arrives at Alice's end. She then applies a unitary $\AC$ that does
\emph{not} depend on $\lm$, for she does not know what Bob is sending, to empty
the channel and arrive at a final state
\begin{equation}
 \ket{\psi_2^\lm} = A\ket{\psi_2^\lm} = \AC \ket{\psi_1^\lm},
\label{eqn6}
\end{equation}
which she can then measure or subject to further processing.

\xb
\outl{Transmission using a number of rounds employing $\BC^\lm_n$,
  etc. }
\xa

This single-round transmission process can be carried out in a number of rounds
in which during the n'th round Bob employs a unitary $\BC_n^\lm$ acting on the
$B+C$ subspace to map an amplitude $c_n\ket{\psi_0}$ into $C$, which is
initially empty, and which travels to Alice, who uses a unitary $\AC_n$ acting
on $A+C$ to remove it from the channel, which is then empty and ready for the
next round. One way to visualize this is that Bob has a domain $B$ of high
dimension, and at the outset splits up the initial amplitude $\ket{\psi_0}$ into
pieces placed in different subspaces of $B$ with the help of a suitable set of
beamsplitters. At round $n$ the unitary $\BC^\lm_n$ interchanges the
appropriate subspace of $B$ with the empty $C$. Alice's $A$ is also large, and
her $\AC_n$ maps whatever Bob has sent into an empty subspace reserved for this
purpose. When the run is completed Alice can then combine the amplitudes in
these different subspaces into a smaller space---e.g., using beamsplitters---or
she can do a similar combination at the end of each round. Of course Alice's
and Bob's unitaries cannot be chosen independently; the two must work together
to design the protocol. What is unknown to Alice is Bob's choice of $\lm$ for a
particular run; this is the information that she can extract at the end.

\xb
\outl{Measurements at intermediate times deferred till the end}
\xa

Some multiple-time protocols employ \emph{measurements} by Alice at
intermediate times. In cases such as the original SLAZ scheme, discussed below
in Sec.~\ref{sct4}, it is possible to store the amplitude that could have
triggered the measuring device in an empty subspace in Alice's domain and put
off the measurement until the protocol is finished. Of course, amplitudes that
correspond to several measurements in succession can be combined, just as in
the case of simultaneous transmission through several channels in parallel, as
discussed earlier.


\section{ Two-way Protocols \label{sct3}}

\subsection{ Gram Matrices \label{sbct3a}}

\xb
\outl{Gram matrices $G^{\mu\nu}(P_j)$ for collection $\{\ket{\psi^\mu}\}$ given
PDI $\{P_j\}$ sum to $G^{\mu\nu}$}
\xa

\xb
\outl{Gram matrices additive over subspaces, invariant under
unitaries}
\xa

Let $\{P_j\}$ be a projective decomposition of the Hilbert space
identity  $I$:
\begin{equation}
 I = \sum_j P_j,\quad P_j=P_j\ad,\quad P_jP_k =\dl_{jk} P_j,
\label{eqn7}
\end{equation}
and let  
$\{\ket{\psi^\mu}\}$, 
$\mu=0,1,\ldots $, be a collection of kets
on the same Hilbert space. The \emph{Gram matrix} 
\begin{equation}
 G^{\mu\nu} = \inpd{\psi^\mu}{\psi^\nu} =\sum_j G^{\mu\nu}(P_j) = 
\sum_j \mted{\psi^\mu}{P_j}{\psi^\nu}
\label{eqn8}
\end{equation}
is \emph{additive} in that it is a sum over contributions from the different
subspaces. In addition, $G^{\mu\nu}$ is \emph{invariant} (or \emph{conserved})
under a unitary operation $U$ that acts on every ket in the collection
$\{\ket{\psi^\mu}\}$. Also, if this unitary acts on only some of the subspaces,
say $P_1$ and $P_2$, and is the identity operator on the others, then while
both $G^{\mu\nu}(P_1)$ and $G^{\mu\nu}(P_2)$ may change, their \emph{sum}
$G^{\mu\nu}(P_1)+G^{\mu\nu}(P_2)$ remains unchanged. That Gram matrices are
additive and conserved plays an important role in what follows.

\xb
\outl{Diagonal elements of G = weights. Off-diagonal elements = overlaps}
\xa

\xb
\outl{Additivity and conservation of overlaps important, but surprising, since
  may refer to runs on different days of the week.}
\xa

\xb
\outl{ Abs(overlap) = fidelity in Qm info. But overlap can be negative}
\xa

We shall refer to the \emph{diagonal} elements $G^{\mu\mu}(P_j)$, which are
non-negative, as \emph{weights}. As these are rather like probabilities, their
additivity and conservation is not surprising. However, that the same is true
of the \emph{nondiagonal} elements $G^{\mu\nu}(P_j)$ with $\mu \neq \nu$,
hereafter referred to as \emph{overlaps}, comes as something of a surprise,
especially since $\ket{\psi^\mu}$ and $\ket{\psi^\nu}$ may refer to two
different runs of an experiment, one on Friday and one on Monday. Nonetheless,
overlaps play a key role in the following analysis, not only as part of the
mathematics but also in a surprising but useful ``intuitive'' way of thinking
about what is going on. The absolute value of an overlap corresponds to a
notion of \emph{fidelity} in quantum information, but in general an overlap
is a complex number, and the fact that it can be negative as well as positive
is a key element in what follows.

\subsection{ Basic Two-Way Protocol \label{sbct3b}}

\xb
\outl{2-way protocol. Alice sends amplitude to Bob who modifies it and returns
  it in successive steps. Final measurement by Alice determines what Bob sent}
\xa

In the following discussion the projective decomposition of the identity
\eqref{eqn7} that will concern us is $\{A,C,B\}$, where
$A$ means that the photon or other quantum particle is
in Alice's domain, $B$ that it is Bob's domain, and $C$ in the channel
connecting them. 
At the beginning of a two-way protocol of the SLAZ type, in which Bob is
sending information to Alice, all of the photon amplitude is in Alice's domain
$A$. She initiates the run by sending some amplitude to Bob through the
channel. He then modifies it and returns some or all of it to Alice, in a
manner that depends on the information $\lm$ he wishes to transmit. Alice
processes what Bob has returned, and begins the second round by again sending
amplitude to Bob, who again returns it, etc. This can go on for $N$ rounds,
following which Alice makes a measurement to determine the value of $\lm$.

\xb
\outl{Details of successive rounds}
\xa

In further detail: At the beginning of round $n$, Alice uses a unitary
$\AC_{n1}$ acting on $A+C$ to map some of the amplitude in $A$ into an empty
channel $C$. This amplitude then flows through the channel to Bob, where he
empties the channel into $B$, does some processing, and then maps some amplitude
back into $C$. This flows to Alice, who empties $C$ into $A$ using a unitary
$\AC_{n2}$. We assume that ``flow through the channel'' does not change
anything, and hence it is convenient not to think of $C$ as divided into
close-to-Alice, close-to-Bob, and in-between subspaces, but simply imagine that
Alice and then Bob and then Alice are acting on a single $C$ subspace. Alice
uses unitaries that act on $A+C$ and are independent of $\lm$, while Bob uses
unitaries $\BC_n^\lm$, that depend on the information $\lm$ he wants to
transmit, which act on $C+B$. Both the Alice and Bob unitaries will in general
depend upon the round $n$, but Alice's do not depend upon $\lm$. In addition we
impose the restriction that Bob's actions are \emph{passive} in the sense
that the magnitude of the amplitude he sends back to Alice in round $n$ cannot
be greater than what he has just received. This last condition clearly
differentiates these two-way protocols from the one-way protocols of
Sec.~\ref{sbct2b}.

\xb
\outl{Intermediate time measurements replaced by unitary operations}
\xa

The requirement that Alice and Bob only employ \emph{unitary} operations
simplifies the analysis. It is true that various published protocols of this
type, including the original SLAZ version to be discussed in Sec.~\ref{sct4},
employ nonunitary measurements at intermediate times. In the cases of interest
to us these measurements can be replaced by unitary operations which allow the
measurements to be put off until the end of the run, in a manner indicated
in Sec.~\ref{sbct2b} and employed in the discussion in Sec.~\ref{sct4}.

\xb
\outl{Costs and total Cost. Cost differs from probability. }
\xa

To quantify the channel usage for these protocols we use the notions of
\emph{Cost}, equal to the absolute square of the amplitude for a single use of
the channel, and \emph{Total Cost} for the sum of the Costs involved in a
single experimental run, as in Sec.~\ref{sbct2a}, see \eqref{eqn2}. An
important issue connected to claims that these protocols are counterfactual has
to do with the difference between Cost and probability, as will be discussed
later for the SLAZ protocol in Sec.~\ref{sct4}---the importance of this has
already been noted in Sec.~\ref{sbct2a}. In particular we will be interested in
identifying protocols that minimize the overall Cost, as in the example
discussed next.

\subsection{ Sending One Classical Bit \label{sbct3c}}

\xb
\outl{Bob transmits single bit $\lm=0$ or $1$. $G^{\mu\nu}(A)$ at
  beginning and end}
\xa

\xb
\outl{Aim: reduce $G^{\mu\nu}(A)$ overlap to zero, preserve 
weights}
\xa

In the simplest SLAZ protocol Bob wants to send a single classical bit, $\lm=0$
or $1$, to Alice. At the start all of the amplitude is in $A$ for both a
$\lm=0$ and a $\lm=1$ run, so all four of the initial Gram matrix elements
$G^{\mu\nu}_0(A)$, where $\mu$ and $\nu$ are the possible values of $\lm$, are equal to $1$. The
goal is that after $N$ rounds the Gram matrix will be
\begin{equation}
 G^{\mu\nu}_N(A) = \dl_{\mu\nu}.
\label{eqn9}
\end{equation}
That is to say, the final result for a $\lm=0$ run is orthogonal to that for a
$\lm=1$ run, and Alice, by making a measurement in an appropriate basis, can
determine which bit $\lm$ Bob sent. Hence the aim of the protocol is that the
\emph{overlaps}, the diagonal elements $G^{01}(A)$ and $G^{10}(A)$, relating
the two different types of run, should decrease from $1$ to $0$, while the
diagonal \emph{weights} $G^{00}(A)$ and $G^{11}(A)$, remain equal to $1$.

\xb
\outl{FULL protocol vs PARTIAL (no signal for $\lm=1$); only FULL considered
  here}
\xa

At this point it is worth noting that if both weights are not maintained---for
example if at the end $G^{00}(A)=1$ while $G^{11}(A)=G^{01}(A)=0$---Alice can
still extract the value of $\lm$ by measuring whether or not the photon is in
the state $\ket{\psi^0}$. Let us call this, for want of a better term, a
\emph{partial} protocol in contrast to a \emph{full} protocol that results in
\eqref{eqn9}. A partial protocol can be used for one-way transmission, and
the obvious advantage is that it costs nothing to transmit $\lm=1$. A possible
disadvantage is that when Alice's measurement reveals nothing it could be
because of some failure in the channel or in the measuring device. In the
present discussion we focus on \emph{full} protocols.

\xb
\outl{Alice sends all to Bob on first step. Cost=2 for any `Qm' info. Two-way
  protocols can have Cost=1 for both classical bits; product not less than 1}
\xa

A very simple way to implement such a protocol is that on the very first step
Alice sends the entire amplitude to Bob, with a Cost of $1$ for this use of the
channel. Bob then simply modifies this using the unitary $\BC^\lm$ and sends it
back to Alice, either in one round or several rounds, with Alice sending
nothing back. The Cost for using the channel in the Bob-to-Alice direction is
also $1$, see the discussion in Sec.~\ref{sbct2b}. Hence a total Cost of $2$
for the protocol as a whole. Notice that since there is no restriction on
$\BC^\lm$ this rather trivial protocol can be used to send ``quantum''
information. From the perspective of Cost, two-way protocols of the kind under
discussion, in which initially all of the amplitude is on Alice's side, are
interesting because a \emph{classical} bit, $\lm=0$ or $1$, can be sent from
Bob to Alice at a total Cost of $1$ rather than $2$. And as shown below in
Sec.~\ref{sbct3d}, the product of the Costs for $\lm=0$ and $1$ cannot be less
than $1$.

\xb
\outl{Notation $\ket{a;c;b}= \ket{a_1,a_2,a_3;c_1,c_2;b_1,b_2}$}
\xa

To discuss the successive steps in protocols that optimize the Cost, we need an
appropriate notation. We will represent kets, thought of as column vectors, in
the way suggested by the following example
\begin{equation}
\ket{\psi} = \ket{a;c;b}= \ket{a_1,a_2,a_3;c_1,c_2;b_1,b_2}
\label{eqn10}
\end{equation}
where the dimensions of the $A$, $B$, and $C$ subspaces are $d(A)=3$, $d(B)=2$
and $d(C)=2$, so $\ket{\psi}$ is an element of a 7-dimensional Hilbert space.
Thus $a_1$, $a_2$, $a_3$, are complex numbers forming a 3-component vector
$a$; similarly $c$ and $b$ are 2-component vectors. If
\begin{equation}
\ket{\hat\psi} = \ket{\hat a;\hat c;\hat b}= 
 \ket{\hat a_1,\hat a_2,\hat a_3;\hat c_1,\hat c_2;\hat b_1,\hat b_2},
\label{eqn11}
\end{equation}
is another vector in the same space, its inner product with $\ket{\psi}$ is
given by
\begin{equation}
 \inpd{\hat\psi}{\psi} = \sum_{j=1}^3 \hat a_j^* a_j +
 \sum_{k=1}^2 \hat c_k^* c_k + \sum_{l=1}^2 \hat b_l^* b_l
\label{eqn12}
\end{equation}

Note that we are dealing with a direct \emph{sum} of subspaces,
$A\oplus B\oplus C$, \emph{not} a tensor product of subsystems. In much of what
follows, $B$ is empty or can be ignored, so $\ket{a;c}$ will suffice; 
this and other minor variants in notation should be self-explanatory.

\xb
\outl{Explicit steps in optimum 2-way protocol using d(A)=2}
\xa

Let us start with an extremely simple one-round full protocol with $d(A)=2$,
$d(C)=1$. It consists of the following steps:
\begin{align}
 &\ket{a_1,a_2;c} = \ket{1,0;0} \ra \ket{1/\st,0;1/\st}
\notag\\
& \Ra \ket{1/\st,0;(-1)^\lm/\st} \ra \ket{1/\st,(-1)^\lm/\st;0},
\label{eqn13}
\end{align}
where $0$ means this amplitude is equal to zero; do not confuse it with the
label $0$ for one of the two orthogonal states of a qubit. Here $\ra$ indicates
the action of a unitary on $A+C$ carried out by Alice, and $\Ra$ a
$\lm$-dependent unitary on $C$ carried out by Bob. The action by Bob could
involve intermediate steps requiring the $B$ subspace, but its net effect is
only to change the contents of $C$, so there is no need to include $B$ in the
discussion.

\xb
\outl{Steps described in words}
\xa

In words: At the outset all of the amplitude is in Alice's $A$, $a_1=1$. She
maps half (in the sense of the absolute square) of it into $C$ and sends it to
Bob, who either sends it back unchanged in order to transmit $\lm=0$, or with
the opposite phase to send $\lm=1$. Alice then empties the channel into the
$a_2$ position, using a unitary on $A+C$ that is independent of $\lm$, as it
simply requires interchanging two subspaces. A final measurement by Alice
determines which of the two orthogonal states is present in $A$, and thus which
bit Bob was sending.

\xb
\outl{Changes in the Gram matrix $G^{\mu\nu}$ during each step}
\xa

Next consider what is happening to the Gram matrices $G^{\mu\nu}(A)$ and
$G^{\mu\nu}(C)$ during the successive steps. In particular, the overlap
$G^{01}(A)$ is equal to $1$ at the outset, and the first step reduces it
to $1/2$ by placing $1/2$ in $C$. Bob's action changes $G^{01}(C)$ from
$+1/2$ to $-1/2$, and this negative contribution to the overlap moves back
into $A$ when Alice empties the channel, leading to the desired $G^{01}(A)=0$.
On the other hand, whereas the weight $G^{00}(A)$ is reduced to $1/2$ during
the first step, Bob's action does not change the sign of $G^{00}(C)$, so in the
final step Alice moves this weight back to its initial value of $1$, and
similarly for $G^{11}(A)$. Thus the goals of a full protocol have been
achieved. 

\xb
\outl{Costs and Cost for $\lm=0$ and $1$.}
\xa

The Costs of using the channel are easily evaluated: $1/2$ for the Alice-to-Bob
step and the same for Bob-to-Alice, for a total Cost of $Q^\lm=1$, the same for
$\lm=0$ and $1$. These satisfy the rigorous lower bound worked out below in
Sec.~\ref{sbct3d}, so this protocol is optimal if one uses total Cost as an
appropriate measure of channel usage.

\xb
\outl{Extension to $N > 1$ rounds: $\ep=1/2N$, or
variable $\ep_n>0$ provided $\sum_n \ep_n = 1/2$}
\xa

This protocol is easily extended to an equally efficient version involving
$N$ rounds, $N$ any positive integer. Let
\begin{equation}
 \ep = 1/2N,
\label{eqn14}
\end{equation}
and for the first, $n=1$, round replace \eqref{eqn13} with
\begin{equation}
 \ket{1,0;0} \ra \ket{\sqrt{1-\ep},0;\sqrt{\ep}\,}
 \Ra \ket{\sqrt{1-\ep},0;(-1)^\lm\sqrt{\ep}\,}
 \ra \ket{\sqrt{1-\ep},(-1)^\lm\sqrt{\ep};0},
\label{eqn15}
\end{equation}
while for round $n+1$,
\begin{align}
 &\ket{\sqrt{1-n\ep},(-1)^\lm\sqrt{n\ep};0} \ra
 \ket{\sqrt{1-(n+1)\ep},(-1)^\lm\sqrt{n\ep};\sqrt{\ep}} 
\notag\\
 &\Ra
 \ket{\sqrt{1-(n+1)\ep},(-1)^\lm\sqrt{n\ep};(-1)^\lm\sqrt{\ep}\,} \ra
 \ket{\sqrt{1-(n+1)\ep},(-1)^\lm\sqrt{(n+1)\ep};0},
\label{eqn16}
\end{align}
where it is straightforward to show that there exists a $\lm$-independent
unitary for the last step. The final result at the end of round $N$ is the same
as in \eqref{eqn13}, the case in which $N=1$, and again the total Cost is
$Q^0=Q^1=1$, independent of $\lm$. One can also let $\ep$ depend on $n$,
thus $\ep_n>0$ for round $n$, subject to the condition
\begin{equation}
 \sum_n \ep_n = 1/2,
\label{eqn17}
\end{equation}
and the Cost is again equal to $1$.

\xb
\outl{Protocols with higher Cost might have advantages. E.g., photon
  polarization.}
\xa

There are other protocols with larger Costs which may have some practical
advantage. Thus rather than a scalar amplitude, Alice might use photon
polarization, say horizontal $H$, which Bob could return as $H$ to send $\lm=0$
or rotate to vertical $V$ to send $\lm=1$. In this case the Costs are
$Q^0=Q^1=2$, so twice that for an optimal one-way protocol. However, there is
now no need to maintain a particular phase relation between what is in Alice's
domain and what is available to Bob during each round. If polarization is
easier to maintain than phase---one leaves that up to the experts---one could
imagine the added Cost being worthwhile if Alice has a large apparatus capable
of generating single photons, while Bob, off on a trip to spy on Eve, needs
only something easily carried in a suitcase.

\xb
\outl{Problem if Bob sends $\lm=1$ by NOT RETURNING amplitude. SLAZ gets around
this by a clever trick}
\xa

The protocol used in SLAZ, in which Bob returns the amplitude for $\lm=0$, but
absorbs it or feeds it to a measuring apparatus for $\lm=1$, looks less
promising. Because the $\lm=1$ weight only moves from Alice to Bob it is
difficult to have $G^{11}(A)=1$ at the end of the protocol. In fact SLAZ,
discussed in Sec.~\ref{sct4}, employs a clever trick (``Zeno effect'') to get
around this problem, albeit at the cost of a large number of rounds to keep the
probability of failure small, and a large channel usage Cost for one of the
bits.

\subsection{ Lower Bound on Costs  \label{sbct3d}}

\xb
\outl{Properties of Gram matrix yields lower bound on 1 bit Cost}
\xa

\xb
\outl{Round $n$: Alice: $\AC_{n1} A\ra C$; Bob acts on $C$; Alice: 
$\AC_{n2} C\ra A$}
\xa

\xb
\outl{Bob's action on $C$ passive: amplitude does not increase on return.
This condition NOT used in deriving the bound}
\xa

The additivity and conservation properties of the Gram matrix $G^{\mu\nu}$
introduced in Sec.~\ref{sbct3a} will now be used to obtain lower bounds on the
total Cost of two-way protocols of the sort exemplified by, but not limited to,
the case of $1$ classical bit discussed above in Sec.~\ref{sbct3c}. Using the
$\ket{a;c}$ notation of \eqref{eqn10}---the $b$ entry is not needed
in the following discussion---round $n$ of an $N$ round protocol consists of
the following steps carried out on $A+C$:
\begin{equation}
 \ket{a^\mu;0}_n\ra \ket{\bar a^\mu;c^\mu}_n
 \Ra \ket{\bar a^\mu;\hat c^\mu}_n
 \ra \ket{a^\mu;0}_{n+1}.
\label{eqn18}
\end{equation}
Here $\mu$ labels the bit which Bob is transmitting during this run. Thus after
Alice uses a unitary $\AC_{n1}$ on $A+C$ to move some amplitude,
$\ket{c^\mu}_n$ into an initially empty channel. Bob applies a unitary
$\BC^\mu_n$ to $C+B$, leading to an amplitude $\ket{\hat c^\mu}_n$---note the
circumflex (hat) added to $c$---in the channel. If Bob's action is passive, as
assumed in Sec.~\ref{sbct3c} (and in the later discussion of SLAZ in
Sec.~\ref{sct4}), one would have
\begin{equation}
 \| \hat c^\mu \|_n \leq \| c^\mu \|_n,
\label{eqn19}
\end{equation}
but this conditions is actually not needed to obtain the general results and
inequalities given below, which thus apply equally to one-way multi-time
transmission. As a final step Alice employs a unitary $\AC_{n2}$ on $A+C$ to
empty the channel by placing its amplitude into $A$. It is important that
Alice's unitaries $\AC_{n1}$ and $\AC_{n2}$, unlike Bob's $\BC^\mu_n$, \emph{do
  not depend upon $\mu$}, which can be different in different runs of the
experiment.

\xb
\outl{Changes in Gram matrix in single round and in all $N$ rounds yields upper
bound \eqref{eqn24} for $|\Dl G^{\mu\nu}(A)|$ }
\xa

The change in the Gram matrix associated with $A$ during round $n$ is given by
\begin{equation}
 G_{n+1}^{\mu\nu}(A) - G_n^{\mu\nu}(A)
 = \inpd{a^\mu}{a^\nu}_{n+1} - \inpd{a^\mu}{a^\nu}_n 
 = \inpd{\hat c^\mu}{\hat c^\nu}_n - \inpd{c^\mu}{c^\nu}_n,
\label{eqn20}
\end{equation}
where $\inpd{a^\mu}{a^\nu}_n$ is the inner product of $\ket{a^\mu}_n$ and
$\ket{a^\nu}_n$. The equality follows from the fact that
$G^{\mu\nu}(A+C)$ is invariant under $\AC_{n1}$ and $\AC_{n2}$,
and additive: $G^{\mu\nu}(A+C)=G^{\mu\nu}(A)+G^{\mu\nu}(C)$.
To discuss the total change during $N$ rounds, $n=1, 2, \ldots N$, it is
convenient to define
\begin{equation}
\ket{C^\mu} := \{\ket{c^\mu}_1, \ket{c^\mu}_2, \ldots \ket{c^\mu}_N\},\quad
  \ket{\hat C^\mu} := \{\ket{\hat c^\mu}_1, \ket{\hat c^\mu}_2, \ldots 
 \ket{\hat c^\mu}_N\}
\label{eqn21}
\end{equation}
 with inner products
\begin{equation}
 \inpd{C^\mu}{C^\nu} = \sum_{n=1} ^N \inpd{c^\mu}{c^\nu}_n,\quad
\inpd{\hat C^\mu}{\hat C^\nu} = 
 \sum_{n=1} ^N \inpd{\hat c^\mu}{\hat c^\nu}_n.
\label{eqn22}
\end{equation}
Summing \eqref{eqn20} over $N$ rounds yields the following formula
\begin{equation}
 \Dl G^{\mu\nu}(A) = G^{\mu\nu}_N(A)-G^{\mu\nu}_0(A) = 
 \inpd{\hat C^\mu}{\hat C^\nu} - \inpd{C^\mu}{C^\nu},
\label{eqn23}
\end{equation}
for the total change in the $A$ portion of the Gram matrix during the full
protocol. This quantity is bounded by
\begin{equation}
 |\Dl G^{\mu\nu}(A)| \leq |\inpd{\hat C^\mu}{\hat C^\nu}| + 
 |\inpd{C^\mu}{C^\nu}| 
 \leq \| \hat C^\mu \| \cdot \| \hat C^\nu \| + \| C^\mu \| \cdot \| C^\nu \|
\label{eqn24}
\end{equation}
using the norm $ \inp{C^\mu} = \| C^\mu \|^2$.

\xb
\outl{Total Costs $K^\mu$, $\hat K^\mu$ for Alice$\ra$Bob, Bob$\ra$Alice;
$Q^\mu=K^\mu +\hat K^\mu$ yield upper bound 
$|\Dl G^{\mu\nu}(A)| \leq \sqrt{Q^\mu Q^\nu}$ }
\xa

Next define the total Cost $K^\mu$ for Alice-to-Bob and $\hat K^\mu$ for
Bob-to-Alice uses of the channel, with $Q^\mu$ their sum:
\begin{equation}
 K^\mu = \inp{C^\mu} = \| C^\mu \|^2 , \quad
 \hat K^\mu = \inp{C^\mu} = \| \hat C^\mu \|^2 , \quad
 Q^\mu = K^\mu + \hat K^\mu.
\label{eqn25}
\end{equation}
Combining \eqref{eqn24} and \eqref{eqn25} gives
\begin{equation}
 |\Dl G^{\mu\nu}(A)| \leq \sqrt{K^\mu K^\nu} 
 + \sqrt{\hat K^\mu \hat K^\nu} \leq \sqrt{Q^\mu Q^\nu}.
\label{eqn26}
\end{equation}
This yields an upper bound
\begin{equation}
 \Dl G^{\mu\mu}(A) \leq Q^\mu
\label{eqn27}
\end{equation}
for a non-negative diagonal weight, and for the off-diagonal overlap:
\begin{equation}
 |\Dl G^{\mu\nu}(A)| \leq \sqrt{Q^\mu Q^\nu}.
\label{eqn28}
\end{equation}

\xb
\outl{Application to 1-bit two-way protocol: $Q^0 Q^1 \geq 1$. Is an equality
  for protocols of Sec.~\ref{sbct3c}. For SLAZ $Q^1$ very small; $Q^0$ very
  large.}
\xa

In the particular case of the 1-bit two-way protocol, Sec.~\ref{sbct3c}, the
aim is to reduce $G^{01}(A)$ from its initial value of $1$ to $0$ after $N$
rounds. Setting $\mu=0$ and $\nu=1$ in \eqref{eqn28}, we see that to achieve
this result it is necessarily the case that the Costs $Q^0$ and $Q^1$ for
sending bits $\lm=0$ and $\lm=1$ must satisfy the condition
\begin{equation}
 Q^0 Q^1 \geq 1.
\label{eqn29}
\end{equation}
This is satisfied as an equality with $Q^0=Q^1=1$ for the specific protocols
discussed in Sec.~\ref{sbct3c}, which shows that they are optimal if total Cost
is used as a measure. For more general protocols there is no reason to expect
that the two Costs will be equal, and in that case if, say, the Cost for
$\lm=1$ is made very small, that for $\lm=0$ must be very large. This is in
fact the case for the original SLAZ protocol, as discussed below in
Sec.~\ref{sct4}, which thus provides an interesting illustration of such a
tradeoff.

\section{ The SLAZ Protocol \label{sct4}}

\subsection{ Description of the Protocol \label{sbct4a}}

\xb
\outl{$M$ outer rounds, each with $N$ inner rounds, $1 \ll M \ll N$}
\xa

The original SLAZ protocol differs from the simpler situation discussed in
Sec.~\ref{sbct3c} in two respects. First, it has a \emph{hierarchical
  structure}: there are a large number $M$ of \emph{outer} rounds or cycles,
each of which consists of a large number $N$ of \emph{inner} rounds or cycles,
and the protocol will succeed with high probability provided
\begin{equation}
 1 \ll M \ll N.
\label{eqn30}
\end{equation}
Second, while Bob sends a bit $\lm=0$ by reflecting the amplitude sent by Alice
back into the channel, for $\lm=1$ he simply empties the channel, which can be
described as a unitary operation in which the $C$ amplitude is placed in Bob's
subspace $B$. In addition, the original SLAZ protocol and some of its
modifications involve measurements at intermediate times, and these will be
replaced in the discussion below by unitary operations in the manner suggested
at the end of Sec.~\ref{sbct2b}.

\xb
\outl{Notation. Capital $A$, $C$, etc. for subspaces; $a_j$, etc. for
  amplitudes}
\xa

We use a notation
\begin{equation}
 \ket{\psi} = \ket{a_1,a_2,a_3,a_4;c;b}
\label{eqn31}
\end{equation}
of the form introduced in \eqref{eqn10}, where the $a_j$ are scalar amplitudes
in Alice's domain $A=A_1+A_2+A_3+A_4$, $c$ is the amplitude the channel $C$,
and $b$ is in Bob's domain $B$. Here capital letters are used to denote
subspaces and the corresponding projectors, while lower case letters indicate
(in general complex) scalar amplitudes. While $A_4$ and $B$ are
one-dimensional, one can also make these larger spaces for reasons that will
appear during the discussion. An abbreviated notation is often convenient:
$\ket{a_2,a_3}$ in the case of a unitary acting on $A_2+A_3$ while all the
other amplitudes remain unchanged.

\xb
\outl{Rotation operators $R(\th)$, $R_M$, $R_N$ by angles $\th_M = \pi/(2M),
\th_N = \pi/(2N)$ }
\xa

\xb
\outl{$(\cos\th_M)^M \approx \exp[-\pi^2/(8 M)]\approx 1-\pi^2/(8 M)\approx 1$}
\xa

Central to the discussion are unitary
operators that represent a rotation by an angle $\th$ on a 2-dimensional space:
\begin{equation}
 R(\th) \ket{\al,\bt} = \ket{\al\cos\th - \bt\sin\th, 
 \al\sin\th+\bt\cos\th }.
\label{eqn32}
\end{equation}
In particular, $R_M$ and $R_N$, defined in terms of small angles, play a
central role:
\begin{equation}
  R_M := R(\th_M),\quad \th_M := \pi/(2M),\quad\quad
 R_N := R(\th_N),\quad \th_N := \pi/(2N).
\label{eqn33}
\end{equation}
Note in particular that 
\begin{equation}
 (R_M)^M = (R_N)^N = R(\pi/2);\quad R(\pi/2)\,\ket{\al,\bt} = \ket{-\bt,\al}.
\label{eqn34}
\end{equation}
In view of the fact that $\th_N$ is a small angle, 
the following approximations turn out to be useful:
\begin{align}
 &\cos\th_N \approx \exp[-\th_N^2/2] = \exp[-\pi^2/(8 N^2)] \approx
 1-\pi^2/(8 N^2);
\notag\\
 & (\cos\th_N)^N \approx \exp[-\pi^2/(8 N)] \approx 1-\pi^2/(8 N) \approx 1,
\label{eqn35}
\end{align}
and similarly if $N$ is replaced by $M$.

\xb
\outl{First step of outer round $m$}
\xa

These approximations are useful for understanding the overall structure
of the protocol, which is the following. At the beginning of outer round $m$,
$1\leq m\leq M$, $R_M$ is applied to $A_1+A_2$ to yield,
\begin{equation}
 \ket{a_1,a_2}^\lm = R_M \ket{\bar a_1,\bar a_2}^\lm,
\label{eqn36}
\end{equation}
where $\bar a_1$ and $\bar a_2$ are the values of these amplitudes at the end
of the previous outer round. In general they depend upon which bit $\lm=0$ or
$1$ is being transmitted, whence the superscript label, even though Alice's
operations do not depend upon $\lm$. The very first outer round $m=1$ begins by
applying \eqref{eqn36} to the starting state \eqref{eqn31} with $a_1=1$ and
all the other amplitudes equal to zero.

\xb
\outl{First step of a given outer round followed by $N$ inner rounds }
\xa

\xb
\outl{Details of inner rounds for $\lm=0$, $1$}
\xa

The initial step \eqref{eqn36}  of outer round $m$ is followed by a
sequence of $N$ inner rounds, each involving the following steps, here
displayed using the type of notation employed in Sec.~\ref{sbct3c}, but now
with reference to the subspace $A_2+A_3+C$. 
\begin{align}
 &\ket{a_2,a_3;c=0} \ra \ket{a'_2,a'_3;c=0} \ra \ket{a'_2,0;a'_3}
\notag\\
& \Ra  \ket{a'_2,0;(1-\lm)a'_3} \ra \ket{a'_2,(1-\lm)a'_3;0},
\label{eqn37}
\end{align}
where
\begin{equation}
 \ket{a'_2,a'_3} = R_N \ket{a_2,a_3}.
\label{eqn38}
\end{equation}
In words, Alice applies the unitary rotation $R_N$, \eqref{eqn33}, to
$A_2+A_3$, and then maps $A_3$ into the empty channel. Next comes Bob's action,
indicated by $\Ra$, to either reflect the amplitude $a'_3$ back into $C$ if he
is sending $\lm=0$, or shift it into his domain $B$, leaving the channel empty
if sending $\lm=1$. Alice, who does not know the value of $\lm$, maps whatever
is in the channel back into $A_3$ by a unitary that simply exchanges the
contents of $A_3$ and $C$, and then begins the next inner round. The result of
$N$ inner rounds in succession is
\begin{equation}
 \ket{a_2,a_3} \ra 
\begin{cases} \ket{0,a_2} \text{ for $\lm=0$, } \\
 \ket{(\cos \th_N)^N a_2,0}\approx \ket{a_2,0} \text{ for $\lm=1$. } 
\end{cases}
\label{eqn39}
\end{equation}
where the $\lm=1$ approximation is justified when $N$ is very large, see
\eqref{eqn35}.

\xb
\outl{Final step completing outer round after the $N$ inner rounds}
\xa

\xb
\outl{Final $\lm=0$, $1$ states at end of $M$ outer rounds}
\xa

Following the $N$ inner rounds Alice completes this outer round by applying a
unitary to $A_3+A_4$ that empties the contents of $A_3$ into $A_4$. For
$\lm=1$, $a_3=0$, \eqref{eqn39}, so this emptying step is trivial, while
for $\lm=0$ it is nontrivial, and plays a significant role in understanding
the true Costs of the protocol. In the original SLAZ protocol this emptying
step is replaced by a measurement, but instead of a measurement one can just as
well let the amplitudes accumulate in $A_4$, which is the perspective used
here. 
At the end of the protocol after completing $M$ outer rounds the final result
is
\begin{align}
 & \lm=0: \ket{ a_1= 1-r_1, a_2=0, a_3=0, a_4=r_4, c=0, b=0 }
\notag\\
 & \lm=1: \ket{ a_1= s_1, a_2=1-s_2, a_3=0, a_4=0, c=0, b=s_b },
\label{eqn40}
\end{align}
where the quantities denoted by $r_j$ and $s_k$ are small corrections, of order 
$1/M$ or $M/N$ If these are ignored, all the amplitude is in $A_1$ for $\lm=0$
or $A_2$ for $\lm=1$, and a simple measurement allows Alice to determine
which bit Bob sent. 

\subsection{ Calculation of Costs and Overlap \label{sbct4b} }

\xb
\outl{Total Cost for $\lm=1$}
\xa

It is fairly straightforward to work out the Costs for the SLAZ protocol using
approximations justified by $1 \ll M \ll N$, and the results are summarized in
Sec.~\ref{sbct4c} below. We begin with the case $\lm=1$. If one ignores small
quantities, the nonzero components of $\ket{\psi}_m$ at the beginning and at
the end of outer round $m$ are
\begin{equation}
 a_1 = \cos(m\th_M),\quad a_2 = \sin(m\th_M),
\label{eqn41}
\end{equation}
and since $M\th_M=\pi/2$, at the end of outer round $M$ the result is
the $\lm=1$ line in \eqref{eqn40}.

The probability that the photon arrives in $B$ during outer round $m$---the
probability that Bob will detect it if he uses a measuring device---is the sum
of the absolute squares of the amplitudes in the channel $C$ in the $N$ inner
rounds, as this is an incoherent process:
\begin{equation}
 N(\sin(m\th_M))^2 (\sin(\th_N))^2 \approx (\pi^2/4) (\sin(m\th_M))^2/N.
\label{eqn42}
\end{equation}
Summing over $m$ gives the total probability
\begin{equation}
 K^1 = Q^1=(\pi^2/8)(M/N).
\label{eqn43}
\end{equation}
that the photon will end up in Bob's domain by the end of the protocol, which
is the same as the total Cost for $\lm=1$.

\xb
\outl{Case $\lm=0$. State at end of outer round $m$. Cost for inner round $n$;
  sum over $n$ for outer round; sum for $M$ outer rounds}
\xa

In the case $\lm=0$, any amplitude placed by Alice in $C$ is immediately
returned by Bob, and at the end of each outer round is emptied into $a_4$, so
that at the end of outer round $m$ the state is
\begin{equation}
 \ket{\psi}_m=\ket{a_1 = (\cos\th_M)^m, a_2=0, a_3=0, a_4, c=0 ,b=0}.
\label{eqn44}
\end{equation}
For $m=M$ this is \eqref{eqn40} with $r_1 = \pi^2/(8M)$. Thus at the
end of the protocol $a_2, a_3, c$ and $b$ are strictly zero. The Cost
associated with inner round $n$---note that the channel is used twice---is
\begin{equation}
 2[\sin\th_M \cdot \sin(n\pi/2N) ]^2.
\label{eqn45}
\end{equation}
Summing over $n$ gives a total of $(\pi^2/4)(N/M^2)$ for each outer round,
and hence for $M$ outer rounds a total Cost of
\begin{equation}
Q^0=(\pi^2/4)(N/M).
\label{eqn46}
\end{equation}

\xb
\outl{Total change of overlap $\Dl G^{01}(A)$}
\xa

To compute the total change in overlap $\Dl G^{01}(A)$, note
that since for $\lm=1$ Bob does not return an amplitude, only the 
$\inpd{C^\mu}{C^\nu}$ term in \eqref{eqn23} contributes. The contribution
for inner round $n$ of outer round $m$ is the product of the factors
\begin{equation}
 [\sin\th_M \sin(n\th_N)] \cdot [\sin(m\th_M)\sin\th_N]
\label{eqn47}
\end{equation}
 corresponding to $\lm=0$ and $1$. Summing them yields
\begin{equation}
 (\sin\th_M\sin\th_N) \sum_{m,n}^{M,N} \sin(m\th_M)\sin(n\th_N) =
 (\pi^2/4MN) (4MN/\pi^2) = 1,
\label{eqn48}
\end{equation}
and hence 
\begin{equation}
 \Dl G^{01}(A)=-1,
\label{eqn49}
\end{equation}
as expected.

\subsection{Discussion of Costs and  Probabilities \label{sbct4c} }

\xb
\outl{Summary of $Q^0$, $Q^1$. Discussion of $\lm=1$.}
\xa

To summarize the results of Sec.~\ref{sbct4b}: The total Costs $Q^0$ and $Q^1$
for $\lm=0$ and $1$ are:
\begin{equation}
 Q^0 = (\pi^2/4)(N/M), \quad Q^1=(\pi^2/8)(M/N), \quad Q^0 Q^1 \approx 3.044.
\quad Q^0/Q^1 = 2N^2/M^2.
\label{eqn50}
\end{equation}
Given that $M \ll N$, $Q^1$ is miniscule, $Q^0$ is enormous, 
while their product is of order $1$, and satisfies the rigorous bound 
\eqref{eqn29}.
The case $\lm=1$ is the easiest to understand. Since Bob does not return the
amplitude put into the channel by Alice, the Bob-to-Alice Cost $\hat K^1$ is
zero. The Alice-to-Bob Cost is $|s_b|^2$ in \eqref{eqn40}, i.e., the
probability that at the very end the photon is in Bob's domain. The physical
reason for this is that the process by which the amplitude gets there is
\emph{incoherent}, no quantum interference, since no amplitude goes back
through the channel. Bob could either accumulate these amplitudes until the end
of the protocol and then measure to see if the photon is in $B$, or carry out a
measurement at the end of each inner round; in either case the probability of
his detecting the photon is $|s_b|^2$ in \eqref{eqn40}. The situation is
analogous to the use of intermediate time measurements in a one-way protocol as
discussed at the end of Sec.~\ref{sbct2b}.

\xb
\outl{Large $\lm=0$ Cost from coherent oscillation of amplitude back and forth
  through channel}
\xa
\xb
\outl{Original SLAZ claim $\lra$ small probability of intermediate time
  detection }
\xa

The enormous Cost $Q^0$ for $\lm=0$ comes about because Bob repeatedly returns
the amplitude sent by Alice in a \emph{coherent} process. While the amplitude
bouncing back and forth through the channel is relatively small, of order
$1/M$, multiplying its absolute square by $2N$, the number of times this
amplitude is is in the channel during each outer round, leads to a Cost of
order $N/M^2$ for each outer round, and hence a total of order $N/M$ for the
complete process.

Clearly the large value of $Q^0$ means the claim that the protocol is
counterfactual cannot be maintained if Cost is used as a criterion for channel
use, so it is worth discussing how the authors of SLAZ reached a different
conclusion. In essence their reasoning was based on the small value of the
amplitude in $A_3$ at the end of an outer round just before it is transferred
to $A_4$, as per the discussion in Sec.~\ref{sbct4a}. The absolute square of
this amplitude is the probability that the corresponding detector $D_3$ in
Fig.~2(b) in the SLAZ paper will be triggered. This amplitude was earlier
oscillating back and forth inside the subspace with projector $S=A_2+A_3+C$,
and hence it is reasonable to assume that if this detector triggers, the
photon was earlier in $S$ during all $N$ inner rounds making up this particular
outer round%
\footnote{This assumption can be justified using Consistent Histories; see
the discussion of measurements in \cite{Grff17b,Grff19b}}. %
As this probability is of order $1/M^2$, the
probability that one of the $D_3$ detectors triggers during the $M$ outer
rounds that make up a given run is of order $1/M$, and hence small.

\xb
\outl{Classical objection: Classical analog: info sent using bouncing ball}
\xa

There are two serious objections to using this small probability to justify the
claim that the protocol is counterfactual: one classical and the other quantum.
Let us start with the former. During a particular outer round the photon
amplitude in a $\lm=0$ run rattles back and forth inside $S$ a total of $N$
times, and in particular it is in $C$ a total of $2N$ times. Consider a
stochastic classical protocol for transmitting information in which most of the
time Alice and Bob exchange no information at all. However, with a small
probability $\ep$ Alice sends a little white ball into the channel leading to
Bob, who colors it green or red and sends it back to Alice to convey one bit of
information. She records the color, paints the ball white, and returns it to
Bob who again colors it to send a second bit, and so forth, for a total of $N$
rounds. The average rate of transmitting information is $N\ep$ bits, and one
cannot simply throw away the factor of $N$ and claim that this protocol is in
some sense `counterfactual'.

\xb
\outl{Qm objection: Prob of being in $S=A_2+A_3+C$ cannot, in presence of Qm
  interference, be equated to individual probs for $A_2$, $A_3$, $C$.}
\xa

The quantum difficulty has to do with what can be inferred from the probability
that the photon was in $S=A_2+A_3+C$ during the inner rounds that make up a
particular outer round. One may be tempted to use classical reasoning and
assume that the probabilities of being in each of the mutually exclusive
regions, $A_2$, $A_3$, and $C$, that combine to make up $S$ are well-defined and
sum to the probability of being in $S$. But in the presence of quantum
interference this sort of reasoning is invalid and leads to paradoxes. See the
discussion of parallel channels in Sec.~\ref{sbct2a}.

\section{ Conclusion \label{sct5}}

\xb
\outl{STUFF RELATED TO ORIGINAL SLAZ}
\xa

\xb
\outl{Extensions of SLAZ, bibliography, criticisms \& replies}
\xa

The original SLAZ proposal has motivated a large number of papers; see the
extensive bibliographies in \cite{HnLR21,Liao22}. Merely trying to summarize
them, much less provide a detailed review, lies outside the scope of the
present paper. Broadly speaking, this literature consists of modifications,
extensions, or improvements of the original SLAZ scheme; along with
criticisms of the claim that these protocols are counterfactual and
replies to such criticisms. It is hoped that the following rather brief
comments will provide some orientation.

\xb
\outl{Extensions: Bob reflects w opposite phase; transmit Qm state; use several
photons}
\xa

Significant extensions of the original SLAZ scheme by the last three members of
the original collaboration include: the use of a phase change rather than
absorption to transmit the $\lm=1$ bit \cite{LAZ14}; a scheme to transmit
quantum states by multiple iterations of the original SLAZ scheme
\cite{LAAZ15}; using many photons in place of a single photon to transmit a
classical bit\cite{Liao22}. These and others are certainly interesting ideas
from the perspective of transmitting quantum information, and worth further
exploration.

\xb
\outl{Ctfl claims invalid as per Sec.~\ref{sct4}: Incorrect use of
  PROBABILITIES in situations where there is Qm interference. COSTS remain
  substantial in asymptotic limit }
\xa

\xb
\outl{SLAZ \& related: interesting, but unsupported claims no aid to serious
  study }
\xa

On the other hand, in these and all other extensions or modifications of SLAZ
this author has examined, the claim that the protocol is ``counterfactual,'' in
the sense that the total use of a quantum channel is negligible in the
asymptotic limit, is subject to the same objections discussed in
Sec.~\ref{sbct4c}: An incorrect use of probabilistic reasoning in a situation
where quantum interference means probabilities cannot be defined, and where
even in a classical situation Cost would be better than probability as a
measure of channel usage. The total Cost remains finite in the asymptotic limit
of a very large number of steps, which means that counterfactual claims should
be dropped. Doing so will aid, not hinder, the serious study of these
interesting quantum schemes for transmitting information.

\xb
\outl{CRITICISMS OF CTFL CLAIMS}
\xa

\xb
\outl{Vd Comment in PRL and Reply. Later Aharonov \& Vd protocol is not ctfl}
\xa

Shortly after the original SLAZ publication, Vaidman published a
Comment\cite{Vdmn14} claiming that in the $\lm=0$ case in which Bob reflects
the amplitude rather than absorbing it, the photon which was later (with high
probability) detected by Alice must at an earlier time have been in the channel
$C$. In their Reply\cite{Slao14} the SLAZ authors pointed out this way of
reasoning about events at an intermediate time in the presence of quantum
interference was invalid, and leads to paradoxes, a position supported by the
analysis in Sec.~\ref{sbct4b} above. However, they then repeated their
original counterfactual claim which itself is based on a defective
understanding of probabilities at an intermediate time. A later and much more
extended criticism of counterfactuality claims by Vaidman \cite{Vdmn15}
suffers from the same difficulty as his earlier Comment.

Some years later Aharonov and Vaidman\cite{AhVd19} claimed to have found a
scheme of the general SLAZ type which is genuinely counterfactual. However,
when measurements or absorption of a photon at intermediate times are replaced
by unitary processes---mapping amplitude into an empty subspace reserved for
this purpose, as discussed in Sec.~\ref{sbct4a}---the inequality in
Sec.~\ref{sbct3d} applies to this case and undermines the counterfactual claim.
The fundamental difficulty with such claims is that the Hilbert space projector
which identifies the position of a particle at some intermediate time does not
commute with the one representing the quantum state evolving unitarily in time.

\xb
\outl{CONTRIBUTIONS OF PRESENT PAPER}
\xa

\xb
\outl{Gram matrices a useful addition to Qm info tools}
\xa

\xb
\outl{Additivity over subspaces, invariance under unitaries $\lra$ conserved
  quantity}
\xa

\xb
\outl{Off-diagonal overlaps, not always positive, key for intuitive
  picture of info transfer. Total $-1$ for Alice precisely right for SLAZ}
\xa

The most significant contributions of the present paper to the analysis of
SLAZ-type protocols is the use of Cost as a measure of channel usage, and the
use of Gram matrices for discussing information transfer at intermediate times
in the presence of quantum interference. In particular, the fact that 
these Gram matrices are additive  over subspaces and invariant (``conserved'')
under unitary time transformations, plays a key part in the discussions in
Sec.~\ref{sct3}. A rather surprising feature is the role of off-diagonal
elements, ``overlaps'', as a type of information measure which, unlike most
such measures, is not in general positive. That it can be negative plays a very
significant part in understanding its intuitive role in information transfer.
That its total change on Alice's side must be $-1$ during the course of a
successful protocol is confirmed for the SLAZ protocol in Sec.~\ref{sbct4b}.

\xb
\outl{FUTURE RESEARCH. OPEN ISSUES}
\xa

\xb
\outl{Gram matrix use needs unitary intermediate time steps, so no measurements}
\xa

\xb
\outl{Open issue: Could intermediate Alice measurements altering later steps
$\ra$ improved algorithm?}
\xa

This use of Gram matrices requires that the intermediate time steps be unitary.
In the case of SLAZ, measurements at intermediate times can be eliminated by
mapping photon amplitude into empty subspaces, and this can be achieved in
certain other cases, e.g., the Aharonov and Vaidman protocol\cite{AhVd19}.
However, it is less clear whether something similar could be done in a case in
which, for example, Alice uses measurements at intermediate times to change
later steps in the protocol in hopes of reducing the total Cost. This author
believes that such an improvement is impossible, because measurements
themselves are quantum processes whose description simply requires a large
enough Hilbert space in Alice's domain \cite{ntk01}. But this has not yet been
demonstrated.

\xb
\outl{What is special about CLASSICAL info? Could Cost for 2-way protocol
be less than 2 for a pair of nonorthogonal Qm states?}
\xa

And what is special about \emph{classical} information? Sending an arbitrary
one-qubit quantum state from Alice to Bob using the 2-way protocol of
Sec.~\ref{sbct3c} could be done with a Cost of 2, which is to say twice that of
simply using a 1-way protocol from Bob to Alice. That this is the minimum seems
likely, but has not been demonstrated. What about a two-way protocol with all
the amplitude starting on Alice's side, with the aim of a perfect transmission
of each of two specified \emph{nonorthogonal} states from Bob to Alice---what
would be the minimum total Cost?

\xb
\outl{SLAZ: Cost big for $\lm=0$, infinitesimal of $\lm=1$, is intriguing }
\xa

An interesting feature of the original SLAZ protocol is the enormous ratio
$2N^2/M^2$, see \eqref{eqn50}, of the Costs to transmit $\lm=0$ and $1$, in
contrast to the relatively simple protocols discussed in Sec.~\ref{sbct3c} for
which the ratio is $1$. Because the success of SLAZ depends upon $N$ being much
larger than $M$, this large ratio presumably has something to do with Bob's not
sending anything back through the channel when $\lm=1$. Might there be some
interesting physical principles, in addition to the Zeno effect, hiding here
and waiting to be explored?

\xb 
\outl{Ideas introduced here could be useful for other problems, such as 3
  or more parties.}
\xa

In conclusion it is hoped that the thinking and tools employed in this paper
will be useful for studying other problems of quantum information at
intermediate times in situations where the careless use of ill-defined
probabilities generates paradoxes rather than physical understanding.
In particular, information transfer among three or more parties, of current
interest in the study of quantum networks, might benefit from the sort of
analysis used here.

\xb
\section*{Acknowledgements}
\xa

The author thanks Scott Cohen for correspondence and an anonymous referee for
suggestions for improving the presentation. He is also grateful to
Carnegie-Mellon University and its Physics Department for continuing support of
his activities as an emeritus faculty member.



\end{document}